\documentclass[12pt]{article}
\usepackage{apjfonts}
\usepackage{graphicx}
\usepackage{sidecap}
\usepackage{floatrow}
\usepackage{hyperref}

\hypersetup{
    colorlinks=true,
    linkcolor=blue,
    filecolor=magenta,      
    urlcolor=blue,
    citecolor=blue,
}

\usepackage{geometry}
\usepackage[utf8]{inputenc}
\usepackage{enumitem,amssymb}
\usepackage{ragged2e}
\newlist{thematic}{itemize}{8}
\setlist[thematic]{label=$\square$}
\usepackage{pifont}

\begin{document}
\raggedright
\setlength{\parindent}{0.4cm}
\huge
\noindent
Astro2020 Science White Paper \linebreak

\noindent 
The First Luminous Quasars and Their Host Galaxies \linebreak
\normalsize

\noindent
 \textbf{Thematic Areas:}    Cosmology and Fundamental Physics;  
         Galaxy Evolution \linebreak

\noindent  
\textbf{Principal Author:}

\noindent 
Name: Xiaohui Fan
 \linebreak						
Institution:  University of Arizona
 \linebreak
Email: fan@as.arizona.edu
 \linebreak
Phone:  520-626-7558
 \linebreak

\noindent  
\textbf{Co-authors:} 
Aaron Barth (UCI), 
Eduardo Ba\~nados (MPIA),
Gisella De Rosa (STScI),
Roberto Decarli (INAF-OAS),
Anna-Christina Eilers (MPIA),
Emanuele Paolo Farina (MPIA \& MPA),
Jenny Greene (Princeton),
Melanie Habouzit (Flatiron CCA), 
Linhua Jiang (PKU),
Hyunsung D. Jun (KIAS),
Anton Koekemoer (STScI),
Sangeeta Malhotra (NASA GSFC), 
Chiara Mazzucchelli (ESO),
Fabio Pacucci (Kapteyn, Yale),
James Rhoads (NASA GSFC), 
Dominik Riechers (Cornell, MPIA),
Jane Rigby (NASA GSFC), 
Yue Shen (UIUC), 
Robert A. Simcoe (MIT), 
Dan Stern (JPL),
Michael A. Strauss (Princeton),
Tommaso Treu (UCLA), 
Bram Venemans (MPIA), 
Marianne Vestergaard (NBI),
Marta Volonteri (IAP),
Fabian Walter (MPIA, NRAO),
Feige Wang (UCSB) and 
Jinyi Yang (Arizona)
  \linebreak

\noindent 
\textbf{Abstract:}
\justify
The discovery of luminous quasars at redshifts up to 7.5 demonstrates the existence of several billion $M_{\odot}$ supermassive black holes (SMBHs) less than a billion years after the Big Bang. They are accompanied by intense star formation in their host galaxies, pinpointing sites of massive galaxy assembly in the early universe, while their absorption spectra reveal an increasing neutral intergalactic medium (IGM) at the epoch of reionization. 
Extrapolating from the rapid evolution of the quasar density at $z=5-7$, we expect that there is only one luminous quasar powered by a billion $M_{\odot}$ SMBH in the entire observable universe at $z\sim 9$.
In the next decade, new wide-field, deep near-infrared (NIR) sky surveys will push the redshift frontier to the first luminous quasars at $z\sim 9 - 10$; the combination with new deep X-ray surveys will probe fainter quasar populations that trace earlier phases of SMBH growth. The identification of these record-breaking quasars, and the measurements of their BH masses and accretion properties require sensitive spectroscopic observations with next generation of ground-based and space telescopes at NIR wavelengths. High-resolution integral-field spectroscopy at NIR, and observations at millimeter and radio wavelengths, will together provide a panchromatic view of the quasar host galaxies and their galactic environment at cosmic dawn, connecting SMBH growth with the rise of the earliest massive galaxies. Systematic surveys and multiwavelength follow-up observations of the earliest luminous quasars will strongly constrain the seeding and growth of the first SMBHs in the universe, and provide the best lines of sight to study the history of reionization.

\thispagestyle{empty}
 
\pagebreak
\setcounter{page}{1}
\setcounter{figure}{0}

\begin{center}
{\bf \Large Introduction}
\end{center}

\justify
As the most luminous non-transient objects that can be observed in the early universe, high-redshift quasars are indispensable tracers of early black hole growth, the formation of the first massive galaxies in the universe, and the history of cosmic reionization.
More than 200 quasars have now been discovered at $z\gtrsim6$ (e.g., Fan et al. 2001;  Willott et al. 2010; Venemans et al. 2015; Jiang et al. 2016;  Ba\~nados et al. 2016, Reed et al. 2017, Matsuoka et al. 2018),  with a handful of objects at $z>7$ (Mortlock et al. 2011, Wang et al. 2018, Matsuoka et al. 2019, Yang et al. 2019), and the highest redshift at $z=7.5$ (Ba\~nados et al. 2018). 
Four main research themes emerge from these discoveries and from the multiwavelength follow-up observations of high-redshift quasars:

$\bullet$ SMBHs with masses up to $10^{10} M_{\odot}$ already existed within the first billion years of cosmic history---this places the strongest constraints on BH formation theory in the early universe.
{\bf How massive are the earliest BH seeds, and how did they grow so rapidly to become the observed high-redshift SMBHs?} 

$\bullet$ The spatial density of luminous quasars declines rapidly towards high redshift; this evolution appears to be accelerating beyond $z\sim5-6$, such that they are the rarest objects detected at high redshift. {\bf Is the epoch of the earliest luminous quasars finally within reach?}

$\bullet$ The host galaxies of high-redshift luminous quasars are intensely forming stars, and show a wide range of masses, kinematics and large-scale environments. {\bf How do early BH growth and galaxy formation influence each other at cosmic dawn?} 

$\bullet$ Hydrogen absorption in high-redshift quasar spectra reveal the end of cosmic reionization at $z\sim 6$,  and suggests that  the IGM  rapidly becomes neutral at $z>7$. {\bf Is the IGM largely neutral by \boldmath{$z\sim 8$}? Is the reionization process uniform or patchy?}

In this White Paper, we focus on future systematic surveys for the earliest luminous quasars at $z\sim 7 - 10$, as well as the observational studies of their BH accretion properties, the evolution of their host galaxies and their large scale environment. These surveys will directly test and constrain theories of SMBH growth and BH/galaxy co-evolution in the early universe. 
A number of other Astro2020 White papers discuss the related topics of seeding massive black holes and probing reionization history using quasar absorption lines.

\begin{center}
{\bf \Large Epoch of the First Luminous Quasars}
\end{center}

\noindent The current decade has featured the following progress in the area of high-redshift quasar surveys:

\noindent
-- the discovery of $z>7$ quasars from near-IR   surveys (Mortlock et al. 2011, Ba\~nados et al. 2018);

\noindent
-- systematic studies of faint quasars at $z\sim 6$ from deep optical  surveys (Matsuoka et al 2018); and

\noindent
-- the assembly of statistical samples of luminous reionization-era quasars at $z>6.5$ from a new generation of wide-field imaging surveys (Wang et al. 2019). 

\begin{figure}[htbp]
\vspace{-3.0cm}
\includegraphics[width=0.62\columnwidth, angle=-90]{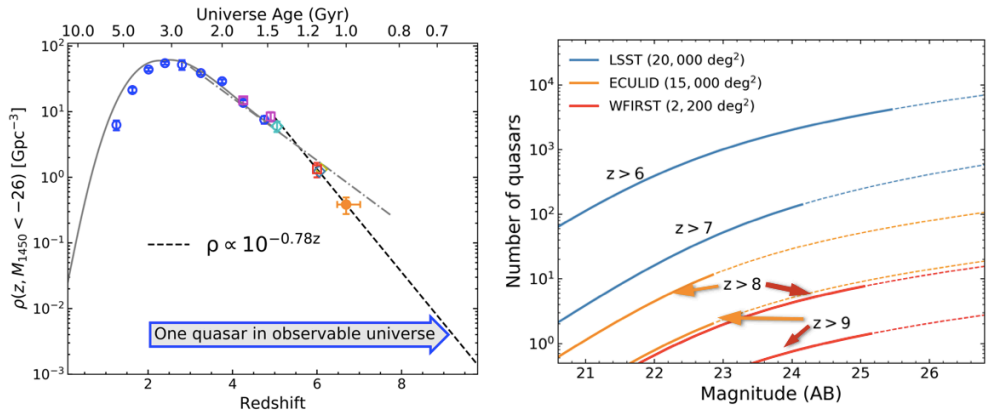}

\vspace{-2.5cm}
\caption{
Left: The density of luminous quasars evolves rapidly in the early universe.  Extrapolating from measurements at $z=5-7$, there will be only $\sim 1$ luminous quasar powered by a billion $M_{\odot}$ BH in the entire observable universe  at $z\gtrsim9$ (Wang et al. 2019). 
Right: Forecast of quasar discoveries in wide-field imaging surveys in the next decade. Dropout selection will discover quasars down to $AB\sim 23 - 25$ (solid lines), establishing a large sample of reionization-era quasars at $z>7$ (LSST), with a few dozen objects at $z>8$, reaching $z\sim 9$ (Euclid, WFIRST). 
}
\end{figure}

These surveys show that while the overall shape of the quasar luminosity function evolves only modestly up to $z\sim 6$ (Matsuoka et al. 2018), its normalization, in other words the overall density of luminous quasars, evolves strongly with redshift (Jiang et al. 2016, Wang et al. 2019).
When modeled as an exponential decline with redshift,
the density of luminous quasars drops by a factor of $\sim 3$ per unit redshift at $z=3-5$; this decline accelerates to a factor of $\sim 6$ per unit redshift at $z=5-7$.
This translates to an e-folding timescale of quasar density evolution of about 400 million years at $z\sim 4$, but as short as 80 million years at $z\sim 7$.  This latter timescale is comparable to the Eddington timescale 
of SMBH growth  of about 45 million years (assuming a radiative efficiency of BH accretion $\sim 0.1$), suggesting that quasar density growth is mainly driven by the maximum rate of accretion onto their central BHs. 
For typical SDSS quasars (UV continuum absolute magnitude of $M_{AB} \sim - 26$, corresponding to $10^9 M_{\odot}$ BH assuming Eddington accretion), their density drops to $<1$~Gpc$^{-3}$ at $z>6$. They are the rarest objects currently known in the early universe.

Extrapolating the density evolution in  the left panel of Figure 1 towards higher redshift, we predict that {\em there will be only $\sim$one $z\sim 9$ quasar at $M_{AB} <- 26$ in the entire observable universe!} 
{\bf This establishes \boldmath{$z=9-10$} as the epoch when the very first luminous quasars appeared in the universe. This redshift range is within reach of survey capabilities in the next decade. Reaching this epoch, as well as establishing large statistical sample of luminous quasars at $z=7-9$, should be a high priority of high-redshift quasar community. }

 \begin{center}
{\bf \Large High-redshift Quasar Surveys in the 2020s}
\end{center}

All luminous quasars at $z>6$ found to date have been selected using wide-field multicolor imaging surveys.
New selection techniques with machine learning (e.g. Schindler et al. 2017)
or Bayesian-based approaches (e.g., Mortlock et al. 2012) have improved the selection efficiency of quasar candidates compared to that of traditional color selection; 
however,  they fundamentally still rely on the  continuum ``drop-out'' caused by IGM absorption blueward of quasar Ly$\alpha$ emission as the main distinguishing feature to separate quasars from low-redshift or galactic contaminants. 
At $z>7$, this Lyman break redshifts into the near-IR wavelengths. The new generation of wide-field near-IR sky surveys, in particular those using LSST, Euclid and WFIRST, will be the key datasets for finding and studying luminous high-redshift quasars in the 2020s.

  \begin{figure}[htbp]
 \vspace{-4.5cm}
\includegraphics[width=0.75\columnwidth, angle=-90]{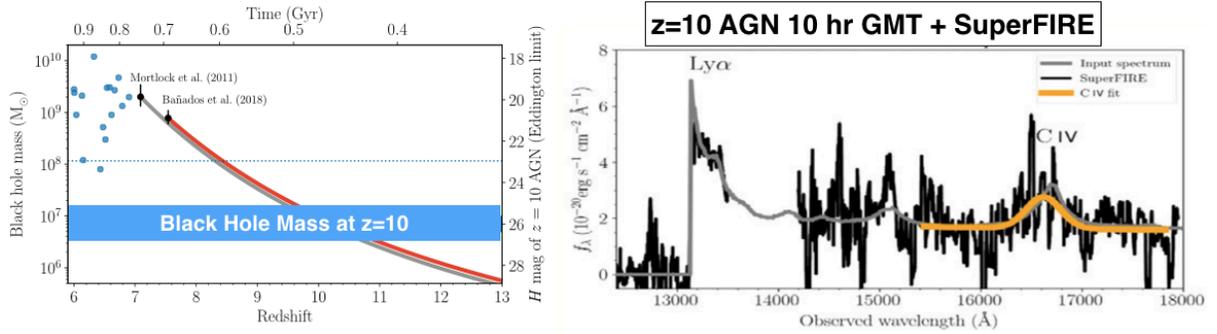}
\vspace{-3.8cm}
\caption{
Left: distribution of SMBH masses in known $z>6$ quasars. Thick lines show their BH masses at higher redshifts assuming Eddington-limited accretion. 
Right: Simulated spectrum of  a faint quasar at $z\sim 10$ powered by a $10^7 M_{\odot}$ BH, with a 10-hour exposure using the Giant Magellan Telescope. Such observations will allow measurements of BH masses and accretion rate in the phase when SMBHs are rapidly growing. Adapted from the \href{https://www.gmto.org/sciencebook2018/}{GMT Science Book}. 
%
}
\end{figure}

Figure 1 (right panel) forecasts the  number of quasars that can be detected in these surveys, based on the quasar luminosity function shape at $z\sim 6$ (Matsuoka et al 2018) and a normalization extrapolating from the measurements at $z\sim 5-7$ (Wang et al 2019). 
For each survey, its limiting magnitude is determined by the depth of the ``dropout'' band below the redshifted Lyman break wavelength. 
For $z\sim6-7$ quasars, the combination of LSST depths and area will yield sample sizes in the hundreds, enabling detailed statistical studies.
Euclid and WFIRST will be the primary source for quasar discoveries at $z>8$, a redshift range in which we expect sample sizes of  a  few dozen. 
Quasar discoveries at $z>9$ will be limited by survey volumes---after all, this is the epoch in which we expect only a small number of luminous quasars to exist within the observable universe. 

A main challenge is the spectroscopic confirmation of $z>8$ quasars selected from these deep near-IR sky surveys. They will be faint, with AB magnitudes of $22 - 25$, requiring observations on large-aperture telescopes.
Past experience suggests that the spectroscopic success rate of spectroscopic identificaiton at the new redshift frontier will be low: Wang et al. (2019) reported a success rate of quasar identification at $z\sim 6.5$ to be $> 30\%$;
current yields at $z>7$ are much lower ($<10\%$).
We expect this efficiency to improve as more quasars are discovered and the size of training sets expanded, 
but the identifications at the highest-redshift frontier will continue to be challenging and highly resource-intensive. 
Identifications of such quasars will be difficult for the current generation of 8-10m telescopes; JWST spectroscopy, while sensitive,  will be more suitable for detailed characterization of already-identified quasars due to its relatively large overhead. High-throughput NIR spectrographs with broad wavelength coverage on the upcoming extremely large telescopes (ELTs; such as E-ELT, TMT, GMT) will be the most effective tools for identification of the earliest luminous quasars (Figure 2, right).
Wide-field IR slitless spectroscopy modes with Euclid and WFIRST will also allow direct confirmation of relatively bright quasars at $z>7$ and the resulting sample will be less affected by color selection biases.


 \begin{center}
{\bf \Large Supermassive Black Hole Growth in the First Billion Years}
\end{center}
 
Quasars discovered at $z>6-7$ are powered by SMBHs up to $10^{10} M_{\odot}$ (Wu et al.\ 2015).
The rapid BH growth in these systems challenges BH formation theory, as it requires either super-Eddington accretion or very massive initial seeds (e.g. Volonteri 2012, Pacucci et al.\ 2017). 
Characterizing their growth requires measuring their SMBH masses and Eddington accretion ratios. 
Masses of the SMBHs in quasars can be estimated by using scaling relations based on the rest UV-frame continuum luminosity and width of the C~IV and Mg~II broad emission lines (Vestergaard \& Osmer 2009, Coatman et al. 2017).  
At $z>7$, these features are redshifted into the NIR band. Current 8-10 m telescopes can only measure BH masses for the most luminous quasars at this redshift.
At higher redshift or lower luminosity, measurements of SMBH masses and accretion rates will require deep spectroscopy with JWST and the ELTs.

Extremely luminous $z\sim 6$ quasars are found to be accreting at close to the Eddington limit (e.g. Willott et al. 2010). 
However,  they are likely to be close to the end of their accretion growth;
studies covering a wide range of luminosities show a diversity of BH accretion rates (e.g., Shen et al. 2019).
Figure 2 (left) shows a compilation of known quasar BH masses at $z>6$, highlighting the two most distant quasars currently known at $z>7$. 
Assuming that all the growth is due to continuous accretion
at the Eddington
limit, their progenitors would be AGN with BH masses of $\sim 10^7$ $M_{\odot}$ at $z\sim 10$,  and  their  observed H-band magnitude would be $\sim 26$.

{\bf Surveys of the more common, lower luminosity  quasars and AGNs at high redshift  will allow studies of 
the earlier phases of the BH accretion process.
These early phases 
are an absolute key to understanding SMBH growth and should be another 
high priority of high-redshift quasar research in the next decade.}
Deep, wide-field X-ray surveys using Athena, AXIS, and Lynx will  reach fainter flux limits compared to optical surveys (Pacucci et al. 2015).
They will also be extremely powerful in uncovering populations of high-redshift quasars that optical/near-IR selection methods are not sensitive to, in particular obscured quasars (which could represent phases of the rapid early BH growth), and in 
 characterizing the BH accretion properties through X-ray emission. However, X-ray surveys will face similar challenges in candidate selection and spectroscopic confirmation, requiring coordinated efforts with deep NIR imaging and spectroscopy on JWST, WFIRST and the ELTs. 

 \begin{SCfigure}
 \vspace{-0.5cm}
\centering
\includegraphics[width=0.35\columnwidth, angle=-90]{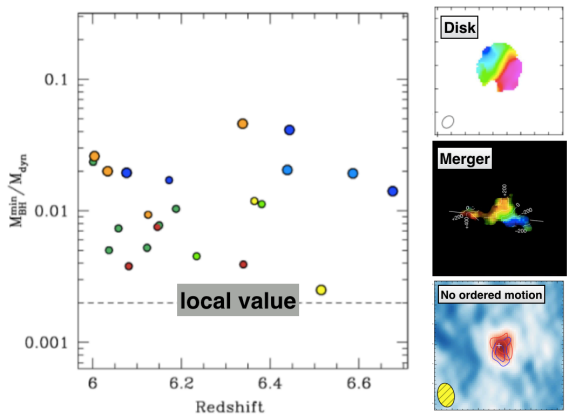}
 \caption{
Left: The SMBH mass to quasar dynamical mass ratio versus redshift for $z > 6$ quasars estimated using ALMA [CII] observations. The ratio observed in local galaxies is marked with a horizontal dashed line. All $z > 6$ quasars clearly lie above the local value by about an order of magnitude, suggesting earlier BH growth compared to host galaxy assembly (Decarli et al. 2018).
Right: ALMA observations show diverse morphologies and kinematics  of $z>6$ quasar hosts. 
}
\end{SCfigure}

\begin{center}
\bf{ \Large BH/Galaxy Co-Evolution at Cosmic Dawn}
\end{center}

High-redshift quasars show highly chemically enriched broad emission lines in their spectra (Hamann et al. 2002), and are among the brightest sources at (sub)millimeter wavelengths (Wang et al.\ 2011, Venemans et al.\ 2018), with a large amount ($\sim 10^{7-8}~M_{\odot}$) of warm dust, as well as abundant  cool molecular gas ($\sim 10^{9-10} M_{\odot}$, Carilli \& Walter 2013) in their host galaxies.
 Assuming that the dust is heated by star formation, this indicates star formation rates of up to a few thousand $M_{\odot}$~yr$^{-1}$, suggesting that rapid BH build-up is coeval with the assembly of massive host galaxies. 
Sensitive, high resolution ALMA observations (Figure 3) reveal a large diversity of host galaxy properties among luminous high-redshift quasars, ranging from isolated, rotationally supported systems, to major mergers with multiple components, to dispersion-dominated compact systems (e.g. Shao et al.\ 2017, Decarli et al.\ 2017). 
The  dynamical mass of the host galaxies, traced by ISM kinematics (mostly from the strong [CII] emission line), appears to be on average about one order of magnitude smaller than the value implied based on the relation between SMBH masses and galaxy masses in the local universe (e.g. Decarli et al. 2018),
suggesting that at these early epochs, the build-up of the most massive BHs occurs {\bf faster} than the assembly of their host galaxies. However, this could be strongly affected by potential biases from selection and by using gas tracers (Volonteri \& Stark 2011, Huang et al.\ 2018).
On a larger spatial scale, while theoretical predictions suggest that luminous  quasars should reside in massive halos in the most overdense environments in the early universe, the observational evidence is  ambiguous at best  (e.g.\ Kim et al.\ 2009, Mazzucchelli et al.\ 2017).
These observations paint a complex picture: luminous high-redshift quasars are among the most active sites of star formation activity, tracing the rapid growth of early massive galaxies; yet their evolutionary sequence, and the interaction between SMBH growth and galaxy assembly is still unknown.  

A key missing observation is the direct detection of stellar light and nebular emission from the high-redshift quasar host galaxies at rest-frame UV to near-IR wavelengths. This would  measure the stellar mass and star formation history, and trace feedback processes in the quasar hosts. The combination of strong nuclear emission from the quasars and low surface brightness of the high-redshift hosts pushes this beyond the capabilities of the current facilities, including  {\em HST}. 
Powerful integral-field spectrographs on JWST and on the ELTs will finally allow such observations. High-fidelity imaging from space, efficient adaptive optics systems for the ELTs, as well as high-contrast imaging techniques developed for exoplanet detection will enable effective removal of the nuclear sources to  map the distribution of stellar light and star-forming regions in  quasar host galaxies.
Deep near-IR imaging (JWST, WFIRST, LUVOIR) will probe faint galaxy  populations and large-scale structure around the earliest quasars.  
Upgrades to ALMA to improve resolution and sensitivity, in concert with the ngVLA at longer wavelengths, would finally enable us to study the ISM properties in the 
quasar hosts down to the scales of individual star-forming regions in a suite of atomic and molecular tracers, to reveal the mechanisms that lead to the build-up of the stellar components and to probe galaxy kinematics to within the sphere of influence of the SMBH.
The union of these facilities will provide a panchromatic  view of the structure of quasar host galaxies and their environments,  connecting SMBH growth with the rise of the earliest massive galaxies.

{\bf Conclusion.} The earliest luminous quasars probe the growth of SMBHs in the early universe and their relation to galaxy formation; they also provide ideal sources for IGM studies at the epoch of reionization.
Quasars, by nature, are multi-wavelength, multi-scale phenomena, and therefore a wide array of new capabilities is required to push quasar investigations toward earlier epochs and fainter populations, and to connect discoveries to physical understanding (Table 1). New facilities in the coming decades will bring into view the epoch of the first luminous quasars in the universe, providing discoveries that will answer the most fundamental questions about the formation and early growth of black holes at cosmic dawn.  
%

%




\begin{table}[h!]
\centering
\caption{Key science questions and required capability mapped to future facilities.}

\smallskip
\begin{tabular}{ccc}
\hline \hline
Science Questions & Critical Observational Needs & Key Future Facilities \\
\hline
Quasar discovery & wide-field surveys & LSST, Euclid, WFIRST\\
 & & Lynx, Athena, AXIS \\
Quasar identification and BH mass & efficient IR spectroscopy & ELTs, JWST \\
Quasar hosts and environment & Multiwavelength IFS & ELTs, JWST, LUVOIR\\
  & & ALMA upgrade, ngVLA\\
 \hline  \hline
\end{tabular}
 \end{table}

\pagebreak
\noindent 
\textbf{References}

\noindent 
Ba\~nados, E. et al. 2016, ApJS, 227, 11\\
Ba\~nados, E. et al. 2018, Nature, 553, 473\\
Carilli, C. L. \& Walter, F.  2013, ARA\&A,  51, 105 \\
Coatman, L. et al. 2017, MNRAS, 465, 2120 \\
Decarli, R. et al. 2017, Nature 545, 457 \\
Decarli, R. et al. 2018, ApJ, 854, 97 \\
Fan, X.  et al. 2001, AJ, 122, 2133 \\
Hamann, F. et al. 2002, ApJ,  564, 592 \\
Huang, K.-W. et al. 2018, MNRAS,  478, 5063\\
Jiang, L.  2016, ApJ, 833, 222 \\
Kim, S. et al. 2009, ApJ, 695, 809 \\
Matsuoka, Y. et al. 2018, ApJ, 869, 150 \\
Matsuoka, Y. et al. 2019, ApJ, 872, L2 \\
Mazzucchelli, C et al. 2017, ApJ, 834, 83 \\
Mortlock, D. et al. 2011, Nature, 474, 616 \\
Mortlock, D. et al. 2012, MNRAS, 419, 390 \\
Pacucci, F. et al. 2015, MNRAS 454, 4 \\ 
Pacucci, F. et al. 2017, ApJ Letters, 850, 2 \\
Reed, S. L. et al. 2017, MNRAS, 468, 4702 \\
Schindler, J. T. et al. 2017, ApJ, 851, 13 \\
Shao, Y. et al. 2017, ApJ, 845, 138 \\
Shen, Y. et al. 2019, ApJ, in press (arXiv: 1809.05584) \\
Venemans, B. P.  et al. 2015, ApJ, 801, L11 \\
Venemans, B. P.  et al. 2018, ApJ, 866, 159 \\
Vestergaard, M. \& Osmer, P. 2009, ApJ, 599, 800 \\
Volonteri, M. 2012, Science 337, 544 \\
Volonteri, M. \& Stark, D. P. 2011, MNRAS, 417, 2085\\
Wang, F. et al. 2018, ApJ, 867, 153 \\
Wang, F. et al. 2019, ApJ, submitted (arXiv: 1810.11926) \\
Wang, R. et al. 2011, AJ, 142, 101 \\
Willott, C. J. et al. 2010, AJ, 139, 906 \\
Wu, X. et al. 2015,  Nature 518, 512 \\
Yang, J. et al. 2019, AJ, submitted (arXiv:1811.11915)

\end{document}